# IQLS: Framework for leveraging Metadata to enable Large Language Model based queries to complex, versatile Data

Sami Azirar
*RWTH Aachen University, Aachen, Germany*

Hossam A. Gabbar*
*Faculty of Engineering and Applied Science, Ontario Tech University, Oshawa, Canada*

Chaouki Regoui
*Digital Technologies, National Research Council Canada, Ottawa, Canada*

*Corresponding author: hossam.gaber@ontariotechu.ca

ABSTRACT: As the amount and complexity of data grows, retrieving it has become a more difficult task that requires greater knowledge and resources. This is especially true for the logistics industry, where new technologies for data collection provide tremendous amounts of interconnected real-time data. The Intelligent Query and Learning System (IQLS) simplifies the process by allowing natural language use to simplify data retrieval . It maps structured data into a framework based on the available metadata and available data models. This framework creates an environment for an agent powered by a Large Language Model. The agent utilizes the hierarchical nature of the data to filter iteratively by making multiple small context-aware decisions instead of one-shot data retrieval. After the Data filtering, the IQLS enables the agent to fulfill tasks given by the user query through interfaces. These interfaces range from multimodal transportation information retrieval to route planning under multiple constraints. The latter lets the agent define a dynamic object, which is determined based on the query parameters. This object represents a driver capable of navigating a road network. The road network is depicted as a graph with attributes based on the data. Using a modified version of the Dijkstra algorithm, the optimal route under the given constraints can be determined. Throughout the entire process, the user maintains the ability to interact and guide the system. The IQLS is showcased in a case study on the Canadian logistics sector, allowing geospatial, visual, tabular and text data to be easily queried semantically in natural language.

## 1 INTRODUCTION

Due to its increasingly complex domain (Sandberg et al. 2022), logistics is a knowledge-intensive field that requires complex and diverse data, which has to be consolidated depending on the use case/application under consideration, such as route planning. The current research mainly focuses on optimizing models to fulfill this task (Borgi, Zoghlami, and Abed 2017), given the required data in a structured format. Retrieving, consolidating and applying the right data is, therefore, a huge task. Relational Databases (RDB), using SQL (Codd 1970), are the industry standard in most domains. Advancements in natural language processing (NLP), as well as the emerging field of process mining (Van Der Aalst 2012), have created the need to use the data more efficiently. Yet, with the increasing complexity of the data, retrieving it with traditional Query Languages requires an increasing level of human expertise, both regarding the database query language and the domain, as well as great effort to create complex and long queries.

Therefore, approaches of natural language queries, which map predefined entities from a pre-built model in the query to entities in the database, have emerged, e.g. in managing instruction texts (Gabbar et al. 2021b, 2021a). Still, one of RDBs limits is that they require complex queries when faced with highly interconnected data. Graph Databases tackle this issue, e.g. SPARQL for knowledge management (Barbieri et al. n.d.). NoSQL databases, such as Neo4j, are therefore used for handling large and diverse datasets, with applications including route planning (Raj et al. 2022). However, they require high levels of effort to set up and maintain since they rely mainly on rule-based reasoning and structured data. Moreover, they need more capability when dealing with a wide variety of data and datatypes (Palovská 2015). Also, RDB's have little flexibility for adapting to new, differently structured data. With the emergence of Large Language Models (LLM), (Vaswani et al. 2017) new possibilities in data and knowledge utilization have emerged, most commonly by querying data with vector databases (Anon [2022] 2023; Guo et al. 2022). However, these approaches deliver one-shot solutions, i.e. they make one thinking step to retrieve the data. This requires increasingly big models while having limited capabilities regarding context awareness when it comes to domain specifics since they are mainly trained on general data. In this paper, the IQLS leverages data schema to tackle this issue. The paper is structured in the following manner: Firstly, it pro-vides a brief overview of recent approaches in complex data querying. Next, it explains the IQLS on a high level and proceeds to describe our case study. After that, the IQLS is explained step by step, followed by a brief analysis of its capabilities and limits. It concludes with directions for future work.

## 2 RELATED WORK

To make use of existing databases without altering them, text-to-SQL-Query approaches have emerged, using semantic trees (Giordani and Moschitti 2010) or knowledge graphs for the mappings (Ahmad, Khan, and Ali 2009). In (Das and Balabantaray 2019), the authors also utilized the database schema to parse correct SQL queries from natural text. Still, one ongoing challenge is the fuzziness of queries, which makes static mapping difficult (Mama and Machkour 2021). Therefore, the emergence of vector embeddings, i.e. embedding the semantic meaning of a text into a latent space, has made more dynamic and versatile text-to-SQL-Query approaches possible by encoding the data into vectors and using the vector similarity to enable semantic parsing of natural texts to SQL Queries (Wang et al. 2021).This is also done by mapping the query vectors directly to the encoded data (Anon n.d.-g). The vector search got further extended with the capabilities of multimodal encoding of complex data into fully vector-based databases (Anon [2022] 2023; Guo et al. 2022). In combination with LLM-based chatbots (Anon n.d.-d; Ansari 2023), and improvements in focused retrievals like Retrieval Augmented Generation (RAG) (Lewis et al. 2021) data querying has become simpler and more accurate.

## 3 III. IQLS

While current methods effectively process unstructured data, they fall short of fully exploiting the knowledge embedded in the metadata of complex and differently standardized but structured data (Becker and Intoyoad 2017). Additionally, there is an absence of a comprehensive framework to iteratively guide Large Language Models (LLMs) through data, incorporating the methodologies of Knowledge Graphs for more efficient data management. Moreover, while frameworks like Langchain (Anon n.d.-f) make it possible to extend LLMs to new data sources easily without alteration, data source selection has to be made by the initial setup, i.e. it requires human effort to include different databases. Common approaches retrieve data in a one-shot process, meaning they directly fetch data in response to a user query. This process requires LLMs to process all data simultaneously, which, despite data schema assistance, can be computationally intensive with numerous databases and is constrained by token length limitations. This enables very little user control over the query process. This is particularly challenging in domain-specific applications like logistics, where such one-shot retrieval may miss crucial domain-specific contexts or in use cases where private hosting is required.

Additionally, the output of these implementations is typically limited to one datatype, most commonly text, which may only partially exploit the data's potential. In this paper, we present IQLS, a framework that leverages the ability of LLMs to be versatile and dynamic, with the structure of other NLP-based approaches leveraging metadata. Instead of one-shot data retrieval, the IQLS guides the LLM stepwise from query understanding through data filtering to the final data retrieval and application. Through this stepwise approach, the IQLS is easily adaptable to new data and is capable of processing complex queries on complex data from various data sources while autonomously selecting the needed ones, thus increasing the retrieved data relevance to the query and minimizing the required computational and human resources required. The IQLS not only retrieves the data needed but also applies the data to the required use case.

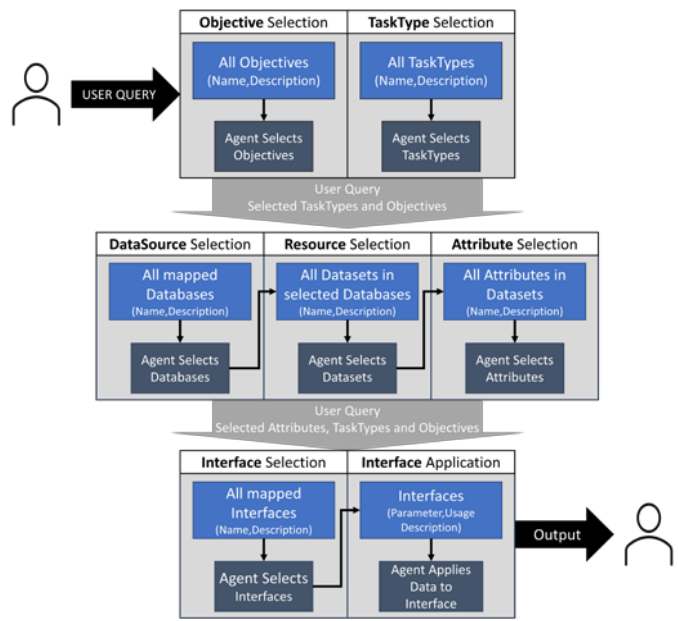

Figure 1 IQLS Framework, agent wraps the LLM

### 3.1 IQLS Process

The IQLS uses an LLM-powered agent. This agent makes decisions based on available choices within an environment connected to a graph database that stores data abstractions. The environment serves as a bridge between the user, the agent, and the actual data. Instead of directly accessing the data, an abstraction is stored in a graph database with different levels of data, descriptions, and metadata.

The process of the IQLS is divided into three main steps, as shown in Figure 1. The first step is to classify the user query based on the *TaskType* and the Objective. Limiting the agent to a predetermined scope of tasks and objectives will set the context to guide the agent toward applicable results. The second step is the data filtering. The agent selects the data sources based on the user query and the abstraction of the data. A *DataSource* represents the highest level of the data hierarchy, such as a database. The environment then searches the Graph Database for the corresponding Resources. A Resource refers to midlevel elements, i.e. the datasets, such as an SQL database table. The agent selects the appropriate resource, after which this process is repeated for the Attributes which correspond to the lower-level elements, like Columns.

In the final step, the interface gets selected and applied. Interfaces are the tools used by the agent to actually retrieve and apply data as the user requires. The interfaces include information retrieval, law retrieval, internal document querying, route monitoring, and route calculation under constraints, as depicted in

### 3.2 System Design

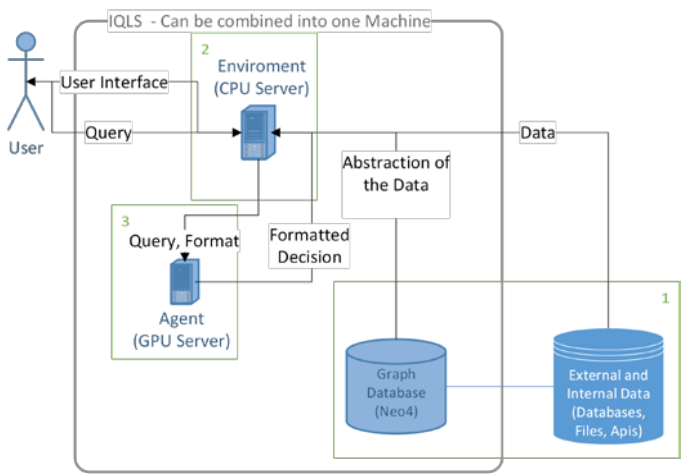

Figure 2 IQLS System Design

The Intelligent Query Language System (IQLS) comprises three key components as illustrated in Figure 2: (1) a graph database for data abstraction, (2) a CPU server for environmental control, and (3) a GPU server for agent operation. Component, the graph database, employs Neo4J (Anon n.d.-h) and stores essential data elements such as labels, descriptions, paths, and links. It is the primary component that requires updates or additions for new data integration. To utilize structured data, users must import it into a Neo4J database conforming to the defined schema. Component 2, which covers all interfaces, manages the user interfaces, and provides methods to interact with various components. It operates using a Python script, a configuration file, and a mapping file. Modifications to this module are necessary only for system improvements or additions, such as updating Neo4J paths or refining agent prompts. It initializes an agent by selecting an LLM and specifying its parameters, then communicates with the agent by delivering a query and the desired output in a Pydantic (Anon n.d.-m) class format. This approach allows for versatility, as any function expressible via a Pydantic class can be implemented. The environment enhances the agent's understanding of queries through context learning, incorporating examples and additional contexts like *TaskTypes* and guidance. Component 3 houses a GPU that drives a large language model. In this paper, the Mixtral 8x7B-Instruct-v0.1 (Jiang et al. 2023) is used. This compatible with all models on Hugging Face (Anon 2024), as long as chat templates are available, and the model supports altering the output logits. Outputs are formatted using the 'lm-format-enforcer' and the Langchain framework (Anon n.d.-i). Fundamental changes to the IQLS System require updates to the agent.

This architecture allows for easy adaptation to new data while minimizing the need for extensive modifications, ensuring stability in system deployment phase. In particular, the agent only accesses data abstractions, making the IQLS suitable for privacy-sensitive applications, with the option to host the agent on the cloud while keeping the data local. For faster inference, the IQLS supports all APIs that follow the OpenAI schema, though this limit altering logits to function calls only, potentially reducing the precision of the filtered data and increasing the error rate. Not all APIs support complex refinements and validators in their function calls. Therefore, the IQLS uses an iterative refinement process for its output. It starts with an initial try, and if the parsing fails, it identifies the attributes that violate constraints or validators. A new Pydantic class is then constructed, consisting solely of the attributes with errors. These are reprocessed by the agent alongside the initial queries and corresponding error messages. If this approach fails, the Langchain *RetryParser* is utilized.

## 4 LOGISTICS CASE STUDY

This paper presents a case study in the Canadian logistics sector, selected for its demand for a straightforward system and the varying complexities in queries and data.

### 4.1 Query Complexity

The IQLS system is designed to handle various complex and diverse queries with ease. Its ability to handle language ambiguity is one of its key features, which enhances user experience by accurately.

| Objective | Description |
| --- | --- |
| Time | Queries about time aspects, like the fastest route. |
| Resources | Queries focusing on resource efficiency, such as the cheapest route. |
| Regulations | Ensuring adherence to local, provincial, and national regulations. |
| Safety | Queries concerning the safety of operations, personnel, and cargo. |
| Distance | Calculating and minimizing travel distances. |
| Certainty | Queries about data reliability, like routes with minimal rain risk. |

Table I Objective Definitions

interpreting natural language queries. Even when dealing with vague expressions or imprecise adjectives like" near," IQLS maintains effective communication with users. Moreover, the system caters to the diverse objectives inherent in logistics-related queries while maintaining the consistency of the underlying data, a key strength of IQLS. This feature enables the system to adapt autonomously to various query goals, eliminating the need for users to navigate between different systems to find answers. The IQLS seamlessly integrates these objectives, ensuring a smooth and efficient user experience. The specific objectives addressed by the IQLS are described in Table I. The final complexity involves task type diversity. Regardless of the objective, queries may target different needs. To address these needs, we defined six task types for the IQLS, as shown in Table II. These task types and objectives determine the scope of the queries the IQLS is developed for.

| *TaskType* | Description |
| --- | --- |
| Information Retrieval | Tasks for finding or gathering data related to a specific objective. |
| Route Planning and Design | Tasks for planning routes based on specific objectives or constraints. |
| Estimations and Forecasting | Tasks for estimating probabilities and forecasting impacts. |
| Supervision & Monitoring | Tasks combining compliance assurance with monitoring updates. |
| Objective Calculation & Evaluation | Tasks for calculating or evaluating metrics for a given objective. |
| Requirement Determination | Tasks for determining requirements to fulfill a specific objective. |

Table II *TaskType* Definitions

### 4.2 *Data*

The IQLS can deal with not only complex queries but also complex data. The data used in the case study was unaltered. However, a prerequisite is that the data must be structured, as this is the main leverage point of the IQLS. This leaves the data complexity in the diversity of data sources, and data formats while being able to be queried semantically across the data types. For this, the case study focuses on four main Data Sources, the first one being the Canadian National Road Network (NRN) Data, which is distributed in thirteen provincial or territorial datasets and includes two linear entities (Road Segment and Ferry Connection Segment) and three punctual entities (Junction, Blocked Passage, Toll Point) [NRN 2023]. These entities are characterized by a range of descriptive attributes, e.g., House Number, Street, Functional Road Class, and Pavement Status. While it is documented, it needs to be clearly interlinked, e.g. the road segments are not natively linked (Anon n.d.-a). However, the validation of the data shows that the geometry borders serve as a sufficient border. The next data source is the 511 API. It includes real-time and static information on Events, Construction, Road Conditions, Transit Hubs, Carpool Lots, Ferries, Service Centres, Travel Info Centres, Truck Rest Areas, Inspection Stations, Roundabouts, Seasonal Loads, Alerts. It can be accessed as an API and returns the data in JSON format. It is available for Ontario as well as Alberta, British Columbia, Yukon, and Prince Edward Island provinces (Anon n.d.-j, Anon n.d.-b, Anon n.d.-c, Anon n.d.-l), but for practical reasons, the case study is limited to the Ontario 511 data source. For regulation data, the CanLII API is utilized, which includes different databases for up-to-date Canadian provincial and federal statutes, regulations, and laws (JSON format). Lastly, this case study included various documents to account for potential internal documents in the application (Anon [2019] 2023).

## 5 IQLS-PROCESSES

### 5.1 *Query Classification and Data Filtering*

First, the user inputs a query to the system. Upon receiving the query, the system retrieves all nodes labelled as *TaskTypes*. Each node contains a name and description of *TaskTypes*, according to the details provided in Table II. Based on the *TaskTypes* present, the system assigns agents to classify the query. While most queries require only one task type, the agent can also utilize multiple task types. Next, the system identifies the objectives based on Table I. The data filtering process begins by retrieving the *DataSources* connected to the selected *TaskTypes*. For each *DataSource*, the agent receives the relevant name and description. Each description contains a summary of the data content, as well as the included Resources and potential applications. If the *TaskType* indicates that route planning is necessary, the NRN Database is also retrieved. The process for the Resources and Attributes is equal. At each step, the lower layer's information is contained in the description. Additionally, at each layer, the user can modify the choices if necessary. This approach allows for filtering large datasets without relying on encoding them into latent spaces, enhancing explainability and user control.

## 5.2 Interfaces

The environment retrieves the possible interfaces and their descriptions based on the chosen *DataSource*, as listed in Table III. The interfaces are not semantically connected to the *DataSources* but based on the data type. For example, only spatial data can be used to plan routes. When the agent needs to fulfill queries that require multiple interfaces, they choose which interfaces to use and in what order. For instance, in route planning, the agent is advised to use all necessary interfaces.

**Information Retrieval:** The Information Retrieval Interface is designed to retrieve various types of data. First, the environment retrieves the path associated with the *DataSources*. The path could be stored directly, with variables for the resources if they are relevant to the path, e.g. if the path is an API endpoint or as a direct link to the *DataSource*. For each chosen resource of the database, the full path will be built if applicable. For all other paths leading to a database or geopackage requiring specific resources, the layer or table will be specified. Finally, for each path, the data will be retrieved and filtered by the chosen Attributes. The Attribute selection is made before data retrieval, if possible, as with gkpg, CSV, or JSON files, otherwise, it's done after loading the data. The data will be converted either to pandas DataFrame or GeoDataFrame (Anon n.d.-k; Jordahl et al. 2019). If the files contain geospatial data, the agent will be asked to decide whether the spatial element is relevant. If it is, the agent also has to specify a location, e.g. Toronto, and a radius of interest. The location is converted to coordinates with the GeoPy module (Anon n.d.-n).

The geospatial data is loaded as a geoDataFrame. If the geometry isn't automatically retrieved, it's converted from the corresponding column, such as from EncodedPolyLine format in the 511 API. Now that the spatial filtering is done, all files will be converted to pandas DataFrames. Some data contains also visual data, e.g. camera recordings. For this, the environment tasks the agent to formulate a question based on the visual input, e.g. asking whether a traffic jam is visible. This question is then processed by the ViLT model (Kim, Son, and Kim 2021), which analyzes the visual input to provide an answer, subsequently stored as a new column mapped to the corresponding row, e.g. camera data. The ViLT model can be hosted on the GPU or on the CPU server, however, with a longer inference time. This will be done for all entries and added as a new column to the table. All tables are stored in a newly initialized SQLite database by the environment, enabling compact, easy, and quick querying. Finally, the SQLite Database will be queried by the agent with respect to the metadata and user query, using the Langchain SQL module.

**Law interface**: The Law Interface is a platform that showcases the use of data sources that have too many attributes or attributes that can't be easily transformed into tabular data. It directly utilizes the CanLII API and goes through the databases and resources. When selecting attributes, the normal approach of choosing from a list is not feasible because there are thousands of potential attributes, such as documents. Instead, a different approach is used, combining retrieval augmented generation with BGE embeddings (Zhang et al. 2023) and (George and Rajan 2022) vector search. This approach can also be CPU-hosted, allowing for a quick selection of attributes while maintaining control over the results and data. The selected regulations are retrieved from the CanLII website, not provided natively by the API. These regulations are then converted into texts and queried by the agent using RAG.

**Internal document retrieval**: The internal document retrieval interface should only be used if the user specifically requests it or if the other interfaces fail to provide the requested information. This system uses RAG in combination with BGFE embeddings and FAISS vector search to query the relevant documents efficiently.

**Route Monitoring Interface**: The Route Monitoring Interface is a tool that can answer questions about specific routes, like finding alerts between points A and B. To do this, the agent first figures out the starting and ending locations that the user is interested in by using the Nominatim module. Next, it imports NRN Road Data as a GeodataFrame and transforms it into a graph using the momepy library (Fleischmann 2019). This graph shows the road segments as directed edges, which reflect the traffic flow direction, and the boundaries of these segments (represented by LineStrings) serve as nodes. The length of each edge is calculated based on the coordinates within the LineString. After dividing the graph into connected components, the largest component is selected for further processing as long as it encompasses over 99.99% of the edges. If not, the system will trigger an error. Once the graph is refined, the agent identifies the nodes closest to the start and end locations. The environment then computes up to ten of the fastest routes, and the chosen road segments are recorded in a database. Finally, the system retrieves additional data in a similar way to the Information Retrieval Interface but without using spatial filtering. Instead, it leverages the Road NID where possible, or it uses the geometry of the road segments to filter information.

| Interface | Description |
|---|---|
| Route Planning Interface | The interface for planning routes under constraints and various objectives. It can be utilized with preselected data to perform various types of route planning. This interface can be used after another interface, if necessary, to perform route planning with the knowledge of the other interface. |
| Law Interface | Includes data about various laws and statutes from different provinces of Canada. It can be used before another interface if needed. This interface does not contain information about specific places or road segments, but only laws and statutes, and should be used solely for this purpose. |
| Specific Information Retrieval Interface | This interface can be used if a specific piece of knowledge is needed. It can be used to retrieve specific data from the previously filtered data, such as the average speed of a road segment, the weather in a certain area, any restrictions on a road segment, something captured on a camera, etc. |
| Road Monitoring | This can be used to check if there are any problems on a route, such as accidents or construction. It can be used after another interface if the user wants to know about potential problems on the route. It can also be used standalone if there is no complex route planning needed and the user wants to know if there could be problems on a route or in a certain area. Additionally, it can be used to monitor a specific area, like a city or a province. |
| Internal Document Retrieval | This can be used if the user specifically needs some internal documents. It can also be used if the required knowledge cannot be found elsewhere. It should be used after the other interfaces have been tried or if the user specifically requests internal documents. |

Table III Interfaces and descriptions

This approach checks for intersections with the route segments, guaranteeing an accuracy of over 10 meters, which is consistent with the NRN Data Specifications.

**Conditional Route Planning**: This interface showcases the applicability of the IQLS and its advantages over standard approaches due to the complexity of the task. This interface, however, is only suited for spatial data; for non-spatial data, the Information Retrieval Interface has to be applied first, e.g. to get a regulation, which then gets applied using the Route Planning Interface. The Data is retrieved following the same approach as the Information Retrieval Interface, only without spatial filtering. When loaded into GeoDataFrames, they are merged into one large GeoDataFrame based on spatial intersections with the road segments. This will be done regardless of the geometry. With this GeoDataFrame, a graph is built with momepy, where all the attributes are edge properties, and the nodes are propertyless. Elements spanning multiple segments are duplicated. To prevent errors, all Nans are filled either with 0 or "None". This graph serves as the model for the road network. To really perform multi-constrained multi-objective path calculation, the agent has to model the pathfinding. For this, the IQLS utilized Pydantic models. The IQLS uses the Driver and the *ConstraintActions* classes. The Driver is modelled through the route network with attributes, while the *ConstraintActions* determine how the Driver interacts with its environment. A modified version of the Dijkstra algorithm is used, as shown Algorithm 1 To determine the best route based on the given constraints and objectives, for each edge for each edge the attribute values of the driver are calculated, skips edges that violate constraints and selects the one with the smallest objective weight. Initially, the Driver has to be determined, so the environment tasks the agent with generating a list of Driver attributes. To execute this algorithm, as shown schematically in Figure 3, initially, the Driver must be determined, so the environment tasks the agent with generating a list of Driver attributes.

All the Driver's attributes should be float values, initialized with 0, i.e. it always should sound up or down, e.g. the time since rest should be counted, not the remaining time until rest. This is important, to simplify the problem towards a minimum problem. The Driver's attributes should be determined based on the user query and a description of each should be generated. The description helps the agent to reason correctly and for later steps to not misunder-

---

**Algorithm 1** Optimized Pathfinding Considering Driver Constraints

1: Let $G = (V, E, w)$ be a weighted directed graph, where $w : E \to \mathbb{R}$.
2: Define driver attributes $D$ and associated state $D_s$.
3: Define constraints set $\mathcal{C}$ and actions set $\mathcal{A}$.
4: Let $s, t \in V$ be the start and target nodes, respectively.
5: Initialize a min-priority queue $Q$ as $\{(0, s, D_s)\}$.
6: Initialize a map Visited $: V \to \mathbb{R} \times V$ with Visited$(s) = (0, \text{null})$.
7: **while** $Q$ is not empty **do**
8:     $(\_, v, D_v) \leftarrow$ dequeue the element from $Q$ with the smallest key.
9:     **if** $v = t$ **then**
10:         **break**
11:     **end if**
12:     **for** each $u \in V$ such that $(v, u) \in E$ **do**
13:         Update $D_v$ according to $\mathcal{A}$ and edge attributes of $(v, u)$.
14:         **if** $D_v$ satisfies $\mathcal{C}$ **then**
15:             Let $\delta = w(v, u) + \text{weight}(D_v)$.
16:             **if** $u \notin$ Visited or $\delta <$ Visited$(u).weight$ **then**
17:                 Visited$(u) \leftarrow (\delta, v)$.
18:                 enqueue $(\delta, u, D_v)$ to $Q$.
19:             **end if**
20:         **end if**
21:     **end for**
22: **end while**

stand. Now, the start and endpoint get determined, as well as the objective attribute, which must be one of the created Driver attributes. The next step involves defining actions to be executed at each step. For this, the agent gets information about which edge attributes are available and their description. Attributes that do not change during each step should have a non-effective action, yet the iterating ensures that for each defined attribute, the agent thinks about its use accordingly. To improve reasoning and enable better later use, each action should be named and described briefly. Each action also includes an operation, e.g. multiply and a Value. The Value could be an Operand object or a string or number. Operand Object defines the attribute of a source, i.e. either the edge or the Driver. The first validator ensures that the source is either the edge or the Driver, and the second validator confirms that the attribute belongs to the source. The action validators ensure that the attribute is indeed the right Driver's attribute. The second action validator ensures that the operation is one of the predefined ones Each action is performed by applying the operator on the Driver's attribute and the Value, e.g. adding the travel time of an edge to the total time of the Driver. Based on the user query, edge and driver attributes, the constraints are defined by the agent. This is a three-stage process. First, a list of constraint names and descriptions is defined by the agent, i.e. the concept of the constraints. This enforces the reasoning of the agent. In the second stage, the constraints are defined by iterating over each name-description pair in the list. For each constraint, the operator, e.g. greater than Operands 1 and 2, gets defined. Operand1 is either a driver or an edge attribute, while Operand2 can also be a text or number. This enables the agent to determine what should be the trigger, e.g. if the travel time without a rest area is greater than 5h or if the road is wet. The action is not defined at this stage, but the name and description, again to reason accordingly and think about the concept first, then the implementation. Finally, the third step determines the action. For the constraints, the action skip edge is always predefined. This skips edges that violate the constraint, e.g. if the travel distance without a fuel station would be large. Again, the operator gets validated. Finally, these get aggregated by the environment into the *ConstraintActions* and the initial Driver object. The *ConstraintActions* objects have the main methods do step, i.e. calculating the new Driver object based on the edge attribute values, and the evaluate condition method, which evaluates all conditions and executes the corresponding actions if the constraints are violated. As shown in 1, this is performed at each edge, always for the next connected edges.

### 5.3 *User Integration*

The user integration in this system is adeptly managed through a Gradio interface (Abid et al. 2019), which is designed to offer two operational modes: automatic and control. In automatic mode, the user inserts a query and the IQLS processes without further interaction. Control mode returns the output of each step to the user, who then corrects and validates the selections. Depending on the query, the output provided to the user varies: for spatial and path data queries, a Folium visualization (Anon n.d.-e) accompanied by textual information is presented, whereas for other types of queries, the output is text-based. An example of Route Monitoring is shown in Figure 4.

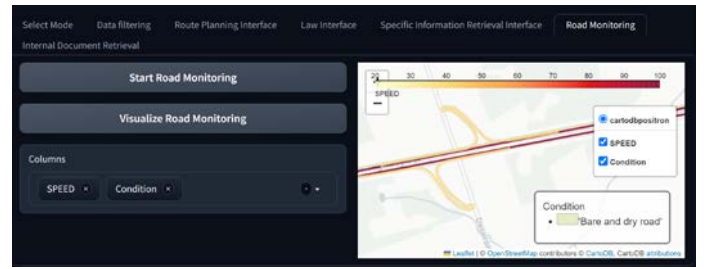

Figure 3 Road Monitoring UI

### 6 EVALUATION

When compared with standard querying methods, the IQLS transforms the task of writing long, complex queries towards making natural language queries and dropdown menu selections. This enables access to complex data for users not familiar with databases. It enables the user to query across databases, independent of the type and origin of data and delivers aggregated results. For example, the query" I am transporting livestock with a truck from Toronto to Ottawa. What do I have to check. I also want to avoid ice on the roads," would require the user to check the internal documents for the checklist. Then, the user would have to get the road weather data and manually construct a route without ice. With the IQLS, the user has only to check the agent's selections and correct them via the Dropdown menu if needed to receive a visualization of the route, as well as the required Information. The IQLS does not require the user to know what is in the available data beforehand, i.e. the user can make queries and then receive the Information and where to find it. In comparison to other LLM-based approaches, the IQLS enables the user to take advantage of the versatility of LLMs while still enabling the user to take control over the querying process in a user-friendly way.

Also, the IQLS can be utilized for any domain and is bounded only by the data structure requirements. Depending on the hardware and query, the local inference time ranges between 5 and 20 minutes with a

40GB Nvidia A100, while API deployment inference takes less than 3 minutes. However, the local deployment enables more control over the output since it enables more precise query results by modifying the length penalty and better output parsing through the output logits. Figure 4. Road Monitoring UI The IQLS extends the capabilities of LLMs to more complex tasks like route planning while still making use of the available data structures. Combined with the output parsing via Pydantic classes, as well as the improved refinement process makes the IQLS versatile in its capabilities. Still, the adaptability of the models strongly depends on the data. Structured data, along with the corresponding metadata, can be inserted naively by importing the data schema in the Neo4J database. More complex APIs, or new database types can be included, yet require programming capabilities. For more complex tasks, like conditional route planning, user involvement increases, for very different use cases. The IQLS enables the user to make queries without using domain-specific language. However, since the IQLS maps the user query to the data based on the schema, the wording of the query also influences the results. For example, the IQLS often included the" Route Number" data when queried for routes since the name is in it.

## 7 FUTURE WORK

To further explore the capabilities and limitations of the IQLS, practical testing in real use environment is necessary. Additionally, the models presented in this paper have yet to be finetuned, leaving room for further improvements. Future research could identify more error-proof ways to parse complex structured outputs from LLMs to enable more versatile applications. Future research can leverage the IQLS Framework in other domains to enable the use of LLM-powered data retrieval on structured data. Furthermore, the integration of ontologies can expand the scope of the IQLS to include unstructured domain-specific data. Finally, we could leverage the IQLS to enable complex use cases on smaller LLMs.

## 8 ACKNOWLEDGEMENT

This project was supported in part by collaborative research and funding from the National Research Council of Canada under the Artificial Intelligence for Logistics Program as well as the NRC-MITACS-RWTH Globalink Research Award Program and by The Smart Energy Systems Lab (SESL) at Ontario Tech University